\documentclass{article}

\usepackage{arxiv}

\usepackage[utf8]{inputenc}
\usepackage[T1]{fontenc}
\usepackage{hyperref}
\usepackage{url}
\usepackage{booktabs}
\usepackage{amsmath}
\usepackage{amssymb}
\usepackage{amsfonts}
\usepackage{array}
\usepackage{multirow}
\usepackage{enumitem}
\usepackage{xcolor}
\usepackage{float}
\usepackage{placeins}
\usepackage{graphicx}
\usepackage{natbib}
\usepackage{tikz}
\usetikzlibrary{arrows.meta, positioning, fit, backgrounds, calc}

\renewcommand{\undertitle}{}
\renewcommand{\headeright}{}

\title{Task-Vector Arithmetic for Emotional Expressivity Control in Language-Model-Based Text-to-Speech}

\author{
  Daniel O. Brito \\
  Instituto de Bioci\^encias, Letras e Ci\^encias Exatas \\
  Universidade Estadual Paulista ``J\'ulio de Mesquita Filho'' (UNESP) \\
  S\~ao Jos\'e do Rio Preto, SP, Brazil \\
  \texttt{daniel.o.brito@unesp.br}
  \And
  Arnaldo Candido Junior \\
  Instituto de Bioci\^encias, Letras e Ci\^encias Exatas \\
  Universidade Estadual Paulista ``J\'ulio de Mesquita Filho'' (UNESP) \\
  S\~ao Jos\'e do Rio Preto, SP, Brazil \\
  \texttt{arnaldo.candido@unesp.br}
}

\renewcommand{\shorttitle}{Task Vectors for Emotion Control in LM-TTS}

\hypersetup{
  pdftitle={Task-Vector Arithmetic for Emotional Expressivity Control in Language-Model-Based Text-to-Speech},
  pdfauthor={Daniel O. Brito and Arnaldo Candido Junior},
  pdfkeywords={emotional text-to-speech, task-vector arithmetic, cross-speaker style transfer, speaker embedding, language-model TTS},
}

\begin{document}
\raggedbottom
\maketitle

\begin{abstract}
This paper investigates whether task vector arithmetic, successful for cross-speaker emotional intensity control in modular text-to-speech (TTS) architectures, transfers to large-scale TTS systems built on language-model backbones with in-context learning (LM-TTS). Through a systematic elimination study over four progressively narrower operands on Qwen3-TTS-12Hz-1.7B---model weights via LoRA fine-tuning, continuous codec embeddings, discrete codec tokens, and the speaker embedding (x-vector) produced by an ECAPA-TDNN encoder jointly trained with the synthesis backbone---we localize the dominant carrier of emotional prosody to the x-vector. Building upon this finding, we propose a training-free method based on centroid arithmetic in x-vector space, $\tau_{\text{emo}} = \mathbb{E}_i[\mathbf{x}(s_i, \text{emo})] - \mathbb{E}_i[\mathbf{x}(s_i, \text{neutral})]$, applied to an unseen target speaker as $\mathbf{x}_{\text{new}} = \mathbf{x}(\text{target}, \text{neutral}) + \alpha\cdot\tau_{\text{emo}}$. Using ESD (English) as the $\tau$ source and emoUERJ (Brazilian Portuguese) as a cross-lingual ground-truth target, we observe average gains of $+0.29$ in emotion2vec cosine over the ICL baseline on English held-out speakers and $+0.09$ on Brazilian Portuguese held-out speakers, while largely preserving identity (WavLM SECS $\gtrsim 0.88$ for the multi-speaker $\tau$ variant) and intelligibility (WER $\approx 0$ in PT-BR). These results offer initial evidence that the dominant carrier of emotional prosody in this class of models is localizable, by elimination, to the co-trained speaker embedding, where training-free centroid arithmetic remains effective even under cross-lingual transfer.
\end{abstract}

\keywords{emotional text-to-speech \and task-vector arithmetic \and cross-speaker style transfer \and speaker embedding \and language-model TTS}

\section{Introduction}

Emotional intensity control in speech synthesis aims to generate speech with desired levels of emotional expressivity, such as ``strong anger'' or ``mild sadness'' \citep{murataSpeakeragnosticEmotionVector2025}. This capability is relevant to applications such as conversational assistants, audiobook narration, and automatic dubbing. The \emph{cross-speaker} variant of the problem seeks to transfer emotional characteristics onto speakers for whom only neutral speech is available, requiring the separation of vocal-identity and emotion representations.

Recently, techniques based on task-vector arithmetic \citep{ilharcoEditingModelsTask2023} have been applied successfully to this problem. Emotion arithmetic consists of obtaining an ``emotion vector'' from the difference between the weights of TTS models fine-tuned on neutral and emotional data of a single speaker \citep{kalyanEmotionArithmeticEmotional2024}. Subsequent work extended the approach by proposing a speaker-agnostic emotion vector, trained over multiple speakers and capable of transferring to speakers unseen during training \citep{murataSpeakeragnosticEmotionVector2025}; others applied the same paradigm to accent \citep{lertpetchpunAccentVectorControllable2026} and to dialect--emotion composition \citep{fengTaskVectorTTS2025}. All these approaches, however, rely on \emph{modular} architectures (e.g., Conformer-FastSpeech2, hereafter CFS2 \citep{guoRecentDevelopmentsESPnet2020}), in which the target attribute has an identifiable functional locus in the weights.

The current speech-synthesis landscape is dominated by language models that operate over discrete speech tokens, spanning both autoregressive (AR) approaches \citep{huQwen3TTSTechnicalReport2026, wangNeuralCodecLanguage2023} and non-autoregressive (NAR) approaches based on discrete diffusion \citep{zhuOmniVoiceOmnilingualZeroShot2026}. These models typically initialize their backbones from pretrained LLMs and employ multi-codebook acoustic tokenizers, representing an architectural class fundamentally distinct from earlier modular approaches. In these architectures, prosody emerges from conditioned autoregressive continuation rather than from parameterized submodules, which raises the question of in which component (backbone, acoustic tokenizer, or speaker encoder) emotional information resides, and of whether task-vector arithmetic transfers to this setting.

In prior work \citep{brito2026sintese}, we investigated explicit emotional-conditioning approaches for Brazilian Portuguese, comparing plain fine-tuning, textual emotion tokens, and the VECL-TTS architecture \citep{gudmalwarVECLTTSVoiceIdentity2024} with dedicated emotion embeddings. The results revealed a competition between emotional control and vocal-identity preservation, with simpler approaches achieving perceptual performance comparable to that of more complex methods. The present work complements that investigation by exploring whether emotional transfer can be achieved implicitly and training-free, dispensing with additional modules or losses, with a focus on LM-TTS.

The contributions of this work are: (1) an investigation that points to the x-vector---the fixed-dimensional speaker embedding extracted by a time-delay neural network (TDNN) trained for speaker verification \citep{snyderXVectorsRobustDNN2018}, here in the ECAPA-TDNN variant \citep{desplanquesECAPATDNN2020}---as the dominant carrier of emotional prosody, and discusses why weight-space arithmetic tends to fail in this class of models; (2) a training-free method for cross-speaker emotional control based on multi-speaker centroid arithmetic in x-vector space, with intensity controllable through the hyperparameter $\alpha$ at inference time and at no additional cost; and (3) a cross-lingual validation with ground-truth reference, in which $\tau$ is extracted from emotional speech in English and applied to unseen Brazilian Portuguese speakers. Code and artifacts for reproduction are publicly available.\footnote{\url{https://github.com/danielbrito91/xvector-emotion-arithmetic}}

\section{Related Work}

\subsection{Task-Vector Arithmetic}

\citet{ilharcoEditingModelsTask2023} formalized task-vector arithmetic: given a pretrained model with weights $\theta_{\text{pre}} \in \mathbb{R}^d$ and a model fine-tuned for a task $t$ with weights $\theta_{\text{ft}}^{(t)}$, the ``task vector'' $\tau_t = \theta_{\text{ft}}^{(t)} - \theta_{\text{pre}}$ can be scaled, summed, and negated to induce controlled behaviors ($\theta_{\text{new}} = \theta + \lambda\cdot\tau$).

\subsection{Emotion Arithmetic in Modular TTS}

\citet{kalyanEmotionArithmeticEmotional2024} were the first to apply task arithmetic to TTS, using a VITS model pretrained on Storynory and fine-tuned by emotion. They demonstrated negation (opposite emotion), addition (composition), and scaling (intensity) operations; a relevant architectural finding is that fine-tuning only the 1D convolutional layers of the text encoder's feedforward block (5M parameters out of 36M total) already produces equivalent emotional speech, indicating that emotion has a specific functional locus in VITS. \citet{murataSpeakeragnosticEmotionVector2025} extended the method to multiple speakers in CFS2 with x-vector conditioning. They diagnosed that the single-speaker $\tau$ of \citet{kalyanEmotionArithmeticEmotional2024} loses speaker consistency in the cross-speaker setting; the proposed solution was to extract $\tau$ from multi-speaker fine-tuning, achieving speaker embedding cosine similarity (SECS), the cosine between the speaker embeddings of the synthesized audio and the reference, comparable to ground truth even for unseen speakers. \citet{lertpetchpunAccentVectorControllable2026} applied the same paradigm to accent in XTTS-v2 with LoRA, while \citet{fengTaskVectorTTS2025} developed the Hierarchical Expressive Vector in F5-TTS for dialect--emotion composition without jointly annotated data.

\subsection{Manipulation of Style and Speaker Embeddings}

A parallel line of work operates on conditioning embeddings rather than on weights. \citet{shaheenExploitingEmotionInformation2023} provided the foundational finding for the present work: speaker embeddings, although designed for identity, contain recoverable prosodic-emotional components. By manipulating only the speaker embedding of an already-trained synthesizer (without retraining it), they raised recognition of the intended emotion from only 9.8\% in the unmanipulated TTS baseline to 57.9\%, direct evidence that emotion resides, latently, in the speaker vector itself---a premise that underpins the x-vector arithmetic proposed below. \citet{joCrossspeakerEmotionTransfer2023} applied manipulation in the latent style space of Tacotron2 (with an adversarial speaker classifier and a cycle-consistency loss), using an SVM hyperplane as the editing direction. \citet{koharaSpeechEmotionControlTexttoSpeech2023} employed per-emotion average x-vectors as conditioning in a VC+VITS pipeline, reaching 96.3\% recognition with ground-truth x-vectors.

More directly related to our operand, \citet{chenEmoKnobEnhanceVoice2024} (EmoKnob) introduced a training-free framework that adds a scaled emotion direction, $\mathbf{u}_s + \alpha\cdot\mathbf{v}_e$ (with $\mathbf{v}_e$ the mean of normalized emotional$-$neutral differences), directly in the speaker-embedding space of a foundation voice-cloning model (MetaVoice-1B), exposing a continuous intensity knob $\alpha$ while preserving identity (evaluated on English only). The arithmetic we adopt is, in its operative form, the same; EmoKnob, however, presupposes the speaker embedding as the operand and conditions on it externally, whereas our setting is an autoregressive LM-TTS with in-context learning whose speaker encoder is co-trained end-to-end with the synthesis backbone.

\citet{suniStyleProsodyControl2025} apply, in a training-free and zero-shot manner for unseen speakers, centroid arithmetic ($\tau = \text{centroid}(\text{emo}) - \text{centroid}(\text{neutral})$, with centroids estimated by the geometric median) over the style embedding of StyleTTS2, but report as a limitation the incompatibility of the method with token-based TTS architectures. Neither line localizes the operand \emph{within} a token-based LM-TTS: \citet{chenEmoKnobEnhanceVoice2024} presuppose the speaker embedding without testing alternative components, and \citet{suniStyleProsodyControl2025} stop at the style embedding of a non-LM synthesizer. We address this empirically through the elimination study of Section~\ref{sec:estudo} and extend the arithmetic to cross-lingual EN$\to$PT-BR transfer (Section~\ref{sec:experimentos}).

\subsection{Emotion Control in Modern LM-TTS}

Recent approaches to emotion in LM-TTS rely on dedicated fine-tuning of substantial cost. \citet{zhouIndexTTS2BreakthroughEmotionally2025} (IndexTTS2) propose an AR LM-TTS with a Text-to-Semantic module, a dedicated perceiver conditioner for emotion, a Gradient Reversal Layer to disentangle identity and emotion, two-stage training, and natural-language instruction via a Qwen3-1.7B fine-tuned with LoRA; the system reaches SOTA on SeedTTS-eval at a cost of three weeks of training on 8 A100 GPUs. \citet{zhangMiniMaxSpeechIntrinsicZeroShot2025} (MiniMax-Speech) propose an AR Transformer with a learnable speaker encoder trained jointly, supporting 32 languages and emotional control via an independent LoRA per emotion category; at the time of its publication (May 2025), the model led the public Artificial Analysis Speech Arena leaderboard. \citet{choEmoSphereEmotionControllableZeroShot2025} (EmoSphere++) employ an adaptive spherical vector in VAD (valence--arousal--dominance) space with Matcha-TTS, reaching an Emotion Classification Accuracy of 93.5\% (seen) and 94.6\% (unseen).

Additionally, \citet{purwarWhenFineTuningFails2026} characterize a relevant failure: LoRA fine-tuning in LM-TTS suffers loss-quality divergence when the data exhibits low acoustic variability. Work with explicit emotional conditioning, such as Qwen3-TTS-CustomVoice, demonstrates that LM-TTS can learn parameterized emotional control when the training objective includes explicit instructions; the Base model (purely ICL), however, does not expose these parameters natively, the context in which our method becomes useful.

\section{Methodology}

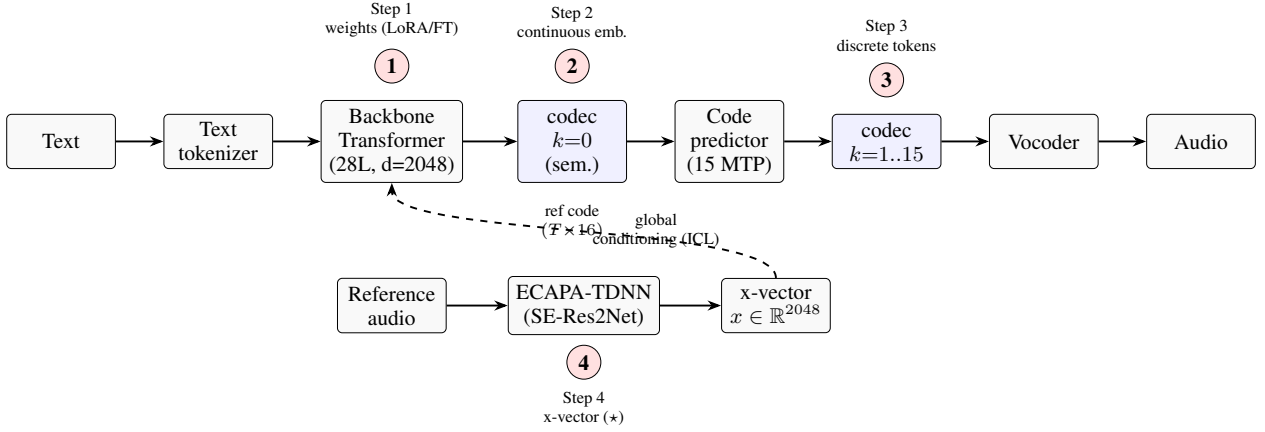
\begin{figure*}[t]
\centering
\resizebox{\textwidth}{!}{%
\begin{tikzpicture}[
  font=\small,
  box/.style={draw, rounded corners=2pt, minimum height=8mm, minimum width=16mm, align=center, fill=gray!5},
  codec/.style={box, fill=blue!6},
  op/.style={circle, draw, fill=red!12, minimum size=5mm, inner sep=0pt, font=\footnotesize\bfseries},
  flow/.style={-{Stealth[length=2mm]}, thick},
  cond/.style={-{Stealth[length=2mm]}, thick, dashed},
  lbl/.style={font=\scriptsize, align=center}
]

\node[box] (text) {Text};
\node[box, right=7mm of text] (ttok) {Text\\tokenizer};
\node[box, right=7mm of ttok] (bb) {Backbone\\Transformer\\(28L, d{=}2048)};
\node[codec, right=8mm of bb] (k0) {codec\\$k{=}0$\\(sem.)};
\node[box, right=7mm of k0] (cp) {Code\\predictor\\(15 MTP)};
\node[codec, right=7mm of cp] (k15) {codec\\$k{=}1..15$};
\node[box, right=7mm of k15] (voc) {Vocoder};
\node[box, right=7mm of voc] (out) {Audio};

\foreach \a/\b in {text/ttok, ttok/bb, bb/k0, k0/cp, cp/k15, k15/voc, voc/out}
  \draw[flow] (\a) -- (\b);

\node[box, below=14mm of bb] (ref) {Reference\\audio};
\node[box, right=9mm of ref] (ecapa) {ECAPA-TDNN\\(SE-Res2Net)};
\node[box, right=9mm of ecapa] (xvec) {x-vector\\$x\in\mathbb{R}^{2048}$};
\draw[flow] (ref) -- (ecapa);
\draw[flow] (ecapa) -- (xvec);
\draw[cond] (xvec.north) .. controls +(up:8mm) and +(down:8mm) .. (bb.south)
  node[midway, right, lbl] {global\\conditioning (ICL)};

\node[lbl, below=2mm of k0] {ref code\\$(T{\times}16)$};

\node[op] (o1) at ($(bb.north)+(0,5mm)$) {1};
\node[lbl, above=0.5mm of o1] {Step 1\\weights (LoRA/FT)};

\node[op] (o2) at ($(k0.north)+(0,5mm)$) {2};
\node[lbl, above=0.5mm of o2] {Step 2\\continuous emb.};

\node[op] (o3) at ($(k15.north)+(0,5mm)$) {3};
\node[lbl, above=0.5mm of o3] {Step 3\\discrete tokens};

\node[op] (o4) at ($(ecapa.south)+(0,-4mm)$) {4};
\node[lbl, below=0.5mm of o4] {Step 4\\x-vector ($\star$)};

\end{tikzpicture}%
}
\caption{Inference pipeline of Qwen3-TTS-12Hz-1.7B-Base and the four candidate operands of the elimination study.}
\label{fig:pipeline}
\end{figure*}

This work evaluates Qwen3-TTS-12Hz-1.7B-Base \citep{huQwen3TTSTechnicalReport2026}, seeking to answer the architectural question of where emotion resides among the components of this model. Unlike frozen ECAPA speaker-verification encoders, the speaker encoder of Qwen3-TTS is \emph{learnable} and co-trained with the backbone under the synthesis objective, aiming at precise identity control; the resulting x-vector is injected directly into the sequence of codec embeddings that conditions the LM. We investigate the steps presented in Figure~\ref{fig:pipeline}, addressed as described in Section~\ref{sec:estudo}.

\subsection{Elimination Study}
\label{sec:estudo}

\subsubsection{Step 1: Task Vectors in Weight Space}
\label{sec:etapa1}

In this step, we investigate whether the procedure of \citet{kalyanEmotionArithmeticEmotional2024, murataSpeakeragnosticEmotionVector2025} ($\tau = \theta_{\text{ft}}^{\text{angry}} - \theta_{\text{ft}}^{\text{neutral}}$) transfers to Qwen3-TTS. To this end, we conduct full fine-tuning (with learning rates $\in \{2\mathrm{e}{-6}, 2\mathrm{e}{-5}\}$) and LoRA (PEFT) with two sets of target modules: (i) attention only (\texttt{q/k/v/o\_proj}, $r=64$, $\alpha_{\text{LoRA}}=128$, $\sim29$M trainable parameters); and (ii) attention plus \texttt{codec\_head} and the 15 \texttt{lm\_head} of the code predictor ($\sim60$M trainable). For each configuration, we sweep learning rates in $\{1\mathrm{e}{-6}, 2\mathrm{e}{-6}, 5\mathrm{e}{-6}, 1\mathrm{e}{-5}, 2\mathrm{e}{-5}, 5\mathrm{e}{-5}, 1\mathrm{e}{-4}\}$ and epochs in $\{4, \ldots, 39\}$ over ESD speaker 0017, with single-speaker/single-emotion data ($\sim$30 min of audio).

\subsubsection{Step 2: Arithmetic on Codec Embeddings}

In this step, we analyze whether emotion is conditioned by the continuous embeddings of the codec tokens. We encode all angry and neutral utterances of 0017 with the tokenizer, look up the 16 embedding tables of the talker, and compute per-codebook centroids: $\text{centroid}_k(\text{emo}) = \mathbb{E}[\mathbf{e}_k(\text{codes})]$. The direction $\tau_k = \text{centroid}_k(\text{angry}) - \text{centroid}_k(\text{neutral})$ is injected in two modes: summed (single sum) and per-layer (independent perturbation per codebook).

\subsubsection{Step 3: Discrete-Token Substitution}

In this step, we analyze whether emotion may reside in the distribution of discrete tokens selected by the codec. Using three parallel pairs from ESD (same text, distinct emotions), we test eight conditions combining codec tokens and x-vector: \texttt{neutral\_baseline}, \texttt{angry\_baseline}, \texttt{full\_swap} (all angry tokens with the neutral x-vector), and five partial variants by codebook subset. The \texttt{full\_swap} condition tests whether emotion is transmitted through the tokens or through the x-vector when both are available in conflicting forms.

\subsubsection{Step 4: Arithmetic on X-Vectors}

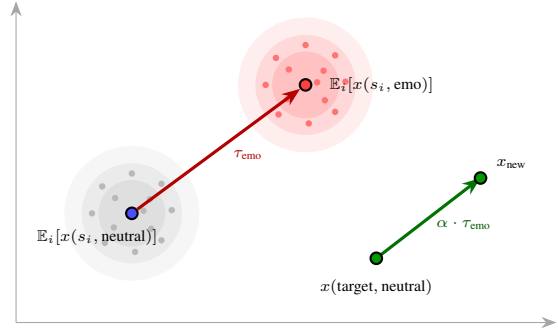
\begin{figure}[H]
\centering
\begin{tikzpicture}[
  font=\small,
  pt/.style={circle, fill, inner sep=1.0pt},
  cen/.style={circle, draw=black, thick, inner sep=1.8pt},
  vec/.style={-{Stealth[length=2.4mm]}, very thick},
  scale=0.85, transform shape
]
\draw[-{Stealth[length=2mm]}, gray!70] (0,0) -- (8.4,0);
\draw[-{Stealth[length=2mm]}, gray!70] (0,0) -- (0,5);

\def\cnx{1.8}\def\cny{1.7}
\fill[gray, opacity=0.07] (\cnx,\cny) circle (1.05);
\fill[gray, opacity=0.09] (\cnx,\cny) circle (0.78);
\fill[gray, opacity=0.11] (\cnx,\cny) circle (0.52);
\foreach \dx/\dy in {0/0.62, 0.46/0.46, 0.62/0.05, 0.46/-0.42, 0.05/-0.6,
                     -0.42/-0.46, -0.62/0, -0.46/0.42, 0.28/0.22, -0.26/0.24,
                     0.3/-0.24, -0.28/-0.22, 0.18/0.04} {
  \node[pt, gray!55] at (\cnx+\dx,\cny+\dy) {};
}
\node[cen, fill=blue!70] (cn) at (\cnx,\cny) {};
\node[below left=0.5mm and -6mm of cn, font=\scriptsize] {$\mathbb{E}_i[x(s_i,\text{neutral})]$};

\def\cex{4.5}\def\cey{3.7}
\fill[red, opacity=0.07] (\cex,\cey) circle (1.05);
\fill[red, opacity=0.09] (\cex,\cey) circle (0.78);
\fill[red, opacity=0.11] (\cex,\cey) circle (0.52);
\foreach \dx/\dy in {0/0.62, 0.46/0.46, 0.62/0.05, 0.46/-0.42, 0.05/-0.6,
                     -0.42/-0.46, -0.62/0,-0.46/0.42, 0.28/0.22, -0.26/0.24,
                     0.3/-0.24, -0.28/-0.22, 0.18/0.04} {
  \node[pt, red!55] at (\cex+\dx,\cey+\dy) {};
}
\node[cen, fill=red!75] (ce) at (\cex,\cey) {};
\node[right=1.5mm of ce, font=\scriptsize] {$\mathbb{E}_i[x(s_i,\text{emo})]$};

\draw[vec, red!70!black] (cn) -- (ce)
  node[pos=0.55, below right=-0.5mm, font=\scriptsize] {$\tau_{\text{emo}}$};

\def\tgtx{5.6}\def\tgty{1.0}
\coordinate (tgt) at (\tgtx,\tgty);
\coordinate (new) at ($(tgt)+(1.62,1.25)$);            
\draw[vec, green!45!black] (tgt) -- (new);
\node[font=\scriptsize, green!35!black] at ($(tgt)!0.5!(new)+(0.55,-0.1)$) {$\alpha \cdot\tau_{\text{emo}}$};

\node[cen, fill=green!60!black] (tgtn) at (tgt) {};
\node[below=1mm of tgtn, font=\scriptsize] {$x(\text{target},\text{neutral})$};
\node[cen, fill=green!60!black] (newn) at (new) {};
\node[above right=-0.5mm and 0.5mm of newn, font=\scriptsize] {$x_{\text{new}}$};

\end{tikzpicture}
\caption{Centroid arithmetic in x-vector space. The emotional direction $\tau_{\text{emo}}$ (difference between the centroids of emotional and neutral x-vectors over $|S|$ source speakers, Eq.~1), applied to an unseen target speaker as $x_{\text{new}} = x(\text{target},\text{neutral}) + \alpha\cdot\tau_{\text{emo}}$, transports the shared emotional axis and preserves the target's identity.}
\label{fig:xvec_arith}
\end{figure}

In this step, we propose the operand $\tau_{\text{emo}}^{(s)} = \mathbf{x}(s, \text{emo}) - \mathbf{x}(s, \text{neutral})$, applied as $\mathbf{x}_{\text{new}} = \mathbf{x}(\text{target}) + \alpha\cdot\tau_{\text{emo}}^{(s)}$ (Figure~\ref{fig:xvec_arith}). Centroids are estimated over 50 utterances per speaker per emotion:
\begin{equation}
\tau_{\text{emo}}^{\text{avg}} = \mathbb{E}_{s \in \mathcal{S}}\big[\mathbf{x}(s, \text{emo})\big] - \mathbb{E}_{s}\big[\mathbf{x}(s, \text{neutral})\big],
\end{equation}
with $\mathcal{S} = \{0011, 0014, 0017, 0020\}$ (the \emph{avg4spk} variant, an operationalization of the principle of \citet{murataSpeakeragnosticEmotionVector2025} in x-vector space rather than in weight space) or $\mathcal{S} = \{0017\}$ (\emph{single0017}). We evaluate three modes:
\begin{enumerate}[label=(\alph*),leftmargin=*,itemsep=0pt]
\item \emph{Same-speaker interpolation}: $\mathbf{x} = (1-\alpha)\cdot\mathbf{x}(0017,\text{neutral}) + \alpha\cdot\mathbf{x}(0017,\text{angry})$;
\item \emph{Cross-speaker interpolation}: same formula, with target-speaker ICL;
\item \emph{Cross-speaker task arithmetic}: $\mathbf{x}_{\text{new}} = \mathbf{x}(\text{target}) + \alpha\cdot\tau$ with target ICL (main mode).
\end{enumerate}

We used $\alpha \in \{0, 0.5, 1, 1.5, 2, 2.5\}$ for EN$\to$EN and $\alpha \in \{0, 1, 1.5, 2, 2.5\}$ for PT-BR.

\subsection{Data}

To obtain $\tau$, we use English data from the Emotional Speech Database (ESD) \citep{zhouEmotionalVoiceConversion2022}, with 350 parallel utterances per emotion per speaker. We adopt four speakers for multi-speaker $\tau$ extraction, namely \{0011, 0014, 0017, 0020\}, balancing gender, with 50 utterances per speaker per emotion. Speaker 0017 alone serves as the single-speaker variant. All recordings are resampled to 24~kHz.

For the cross-speaker English target (held-out), we use two unseen ESD speakers, \{0013, 0019\}, which correspond to the unseen split used by \citet{choEmoSphereEmotionControllableZeroShot2025}, allowing direct comparison with the dimensional SOTA. The evaluation comprises $n=30$ sentences from the official test split ($\text{local\_idx} \in [321,350]$).

For the cross-lingual Brazilian Portuguese target, three speakers from emoUERJ \citep{germanoEmoUERJEmotionalSpeech2021} were used: m03 and m04, male, with parallel pairs; w04, female, without complete parallel pairs, evaluated in a cross-text design. This parallel structure is essential for the falsification test described in Section~\ref{sec:metricas}. The original emoUERJ (44.1~kHz stereo) is resampled to 24~kHz mono with $-1$~dB of headroom to avoid clipping.

\subsection{Metrics, Ceiling, and Baseline}
\label{sec:metricas}

Following the reference framework \citep{choEmoSphereEmotionControllableZeroShot2025}, we adopt the following metrics:

\begin{itemize}
    \item \emph{Emotion embedding cosine similarity} (EECS): cosine between the \texttt{emotion2vec\_plus\_large} embedding of the synthesized audio and that of the paired emotional ground truth (same speaker, same text).
    \item SECS with a WavLM encoder (SECS$_W$): cosine between the WavLM x-vector (\texttt{microsoft/wavlm-base-plus-sv}, an encoder independent of the operand) of the synthesized audio and that of the paired neutral reference. We adopt an encoder external to the TTS pipeline to avoid operand$\leftrightarrow$metric circularity.
    \item \emph{Word Error Rate} (WER): obtained with Whisper-large-v3 \citep{radford2022robustspeechrecognitionlargescale}, configured for the output language and normalized with \texttt{EnglishTextNormalizer} (English) or \texttt{BasicTextNormalizer} (Portuguese).
    \item Naturalness: UTMOSv2 \emph{fusion\_stage3} \citep{baba2024t05voicemoschallenge2024}, winner of the VoiceMOS Challenge 2024. We note that UTMOSv2 is not trained on PT-BR and therefore acts as a relative within-experiment proxy.
\end{itemize}

In every table, we report a natural-ceiling row, computing the same metrics over the real ESD/emoUERJ recordings. For EECS, since $\cos(\mathbf{e}(\text{GT}), \mathbf{e}(\text{GT})) = 1$ is trivial, we use $\text{EECS}_{\text{self}}$ as the natural within-class variance.

The parallel structure of emoUERJ enables a unique metric: $\text{xvec\_cos\_GT} = \cos\big(\mathbf{x}(\text{synth}_{\alpha}), \mathbf{x}(S, \text{emo}, T)_{\text{GT}}\big)$, where $\mathbf{x}$ is the ECAPA-TDNN x-vector. This metric directly tests the arithmetic hypothesis: if $\tau_{\text{emo}}$ captures the emotional direction in a language-agnostic way, then $\mathbf{x}(S, \text{neutral}) + \alpha^{*}\cdot\tau_{\text{emo}}$ should approach $\mathbf{x}(S, \text{emo})_{\text{GT}}$ for some $\alpha^{*}$.

The condition $\alpha = 0$, in all sweeps, corresponds to Qwen3-TTS without any manipulation. Reporting it separately follows the methodology of \citet{shaheenExploitingEmotionInformation2023}, which isolated the effect of speaker-embedding manipulation against the unedited baseline.

\FloatBarrier
\section{Experiments and Results}
\label{sec:experimentos}

\subsection{Elimination Study: Localizing the Emotion Carrier}

Steps 1--3 progressively eliminate the candidate operands of Figure~\ref{fig:pipeline}; Table~\ref{tab:elim} summarizes the outcomes, detailed below.

\begin{table}[H]
\centering
\caption{Summary of Steps 1--3 of the elimination study (perceptual evaluation; full grids, per-codebook norms $\|\tau_k\|$, and the eight substitution conditions in the supplementary notes). Among the operands and regimes tested, none preceding the x-vector yielded emotional control via linear arithmetic.}
\label{tab:elim}
\footnotesize
\begin{tabular}{@{}c l p{3.6cm} p{5.2cm}@{}}
\toprule
Step & Operand & Intervention & Outcome \\
\midrule
1 & Weights (backbone) & FT / LoRA ($\theta^{\text{angry}}_{\text{ft}}-\theta^{\text{neutral}}_{\text{ft}}$) & No useful window: noise (high lr) or calm speech (low lr); off-task laughter with \texttt{codec\_head}. \\
2 & Codec emb. & per-codebook $\tau_k$ (summed/per-layer) & Abrupt ``no effect'' $\to$ degenerate noise transition; no emotional regime. \\
3 & Discrete tokens & \texttt{full\_swap} (angry tokens + neutral x-vec) & \textbf{Calm, coherent output}: the LM ignores token emotion (controlled dissociation). \\
\midrule
\textbf{4} & \textbf{x-vector (ECAPA)} & \textbf{centroid $\tau_{\text{emo}}$ (Eq.~1)} & \textbf{Carrier localized ($\S$\ref{sec:tau_stats}).} \\
\bottomrule
\end{tabular}
\end{table}

\textbf{Step 1 (weights).} Full fine-tuning produces only two regimes: noise (high learning rate) or calm speech indistinguishable from the base model (low learning rate); no intermediate regime that yields intelligible emotional speech is observed. LoRA with the inclusion of \texttt{codec\_head} and the 15 \texttt{lm\_head} of the code predictor produces generic expressivity (with the emergence of laughter in several configurations), without controlled directionality toward the intended emotion. This pattern is consistent with \citet{purwarWhenFineTuningFails2026}. We conclude that emotion was not shown to be controllably capturable through backbone weights in this regime (single-speaker/single-emotion), unlike what is observed in CFS2 \citep{murataSpeakeragnosticEmotionVector2025} and VITS \citep{kalyanEmotionArithmeticEmotional2024}.

\textbf{Step 2 (codec embeddings).} Codebook~1 carries the largest $\tau$ norm ($\|\tau_1\|_2 = 0.137$ vs.\ $\|\tau_{\text{summed}}\|_2 = 0.293$), consistent with its reported role in RVQ codecs of encoding fine acoustic features. However, extensive $\alpha$ sweeps in the summed and per-layer modes produce abrupt transitions from ``no effect'' to ``degenerate noise'', without any intermediate regime in which intelligible emotional speech emerges. The interpretation is that adding a continuous vector to the embeddings of discrete tokens shifts the input toward regions of the space that correspond to no valid token---that is, out of the training distribution of the backbone (out-of-distribution). Emotion, if it resides in this layer, is likely encoded in \emph{which} tokens are selected (a discrete distribution), and not in a continuous direction summable to each token's embedding; it is therefore not accessible via linear arithmetic in our sweeps.

\textbf{Step 3 (discrete tokens).} This step rests on a controlled-dissociation experiment, which offers the most direct evidence of the study. The \texttt{full\_swap} condition, in which all codec tokens of the angry utterance are combined with the neutral x-vector, produces calm and coherent speech, indistinguishable from the neutral baseline, even though the \texttt{ref\_code} carries the entire acoustic signal of angry. The inverse condition (\texttt{angry\_baseline}) sounds angry. This dissociation points to the x-vector as the dominant carrier of emotion: the LM appears to ignore the emotional coloring of the codec tokens in favor of the direction indicated by the x-vector. The partial substitutions by codebook subsets produce degeneration (internally inconsistent \texttt{ref\_code}, with codebooks merged from two utterances, outside the training distribution), which corroborates that coherence depends on the internal consistency of the token set.

\subsection{Step 4: Arithmetic on X-Vectors}
\label{sec:tau_stats}

The geometry of the emotional directions (norms and pairwise cosines, computed over 50 utterances/speaker/emotion; full values in the supplementary notes) supports the method. The multi-speaker average (\emph{avg4spk}) attenuates the $\tau$ norm by 36--60\% relative to \emph{single0017}, since the speaker-conditional component is canceled. For speaker 0017 (angry), we have $\|\mathbf{x}(\text{neutral})\| = 16.70$, $\|\mathbf{x}(\text{angry})\| = 16.89$, and $\cos(\text{neutral}, \text{angry}) = 0.988$: emotion is a small perturbation in norm ($\|\tau\|/\|\mathbf{x}(\text{neutral})\| = 0.154$), yet effective in direction.

The emotional directions are not strictly orthogonal. The pairwise cosines lie in $[0.35, 0.72]$ for \emph{single0017} and decrease to $[0.21, 0.61]$ for \emph{avg4spk}. The multi-speaker average substantially decorrelates the directions, especially $\cos(\tau_{\text{happy}},\tau_{\text{sad}})$ ($0.58 \to 0.24$), which is consistent with the hypothesis of \citet{murataSpeakeragnosticEmotionVector2025} and relevant to the linear composition of multiple $\tau$ noted in future work.

\FloatBarrier
\subsubsection{Cross-Speaker Task Arithmetic (EN held-out)}

\begin{table}[H]
\centering
\caption{Objective EN held-out results at $\alpha^{*}_{\text{emo}}$ (mean over the unseen targets \{0013, 0019\}, $n=30$ per combination). Ceiling = paired human recording (natural reference band, not a strict upper bound; for SECS$_W$ it reflects the human's own identity shift under emotion, hence the near-neutral baseline can exceed it). Base = pure ICL ($\alpha=0$). $\uparrow$/$\downarrow$: higher/lower is better. Bold: best proposed variant per metric per emotion.}
\label{tab:en_operating}
\footnotesize
\setlength{\tabcolsep}{5pt}
\begin{tabular}{l l cccc}
\toprule
Emotion & System & EECS\,$\uparrow$ & SECS$_W$\,$\uparrow$ & UTMOS\,$\uparrow$ & WER$_{\text{norm}}$\,$\downarrow$ \\
\midrule
\multirow{4}{*}{Angry}
 & Ceiling    & 0.957 & 0.898 & 3.048 & 0.069 \\
 & Base       & 0.539 & 0.945 & 3.378 & 0.061 \\
 & avg4spk    & \textbf{0.925} & \textbf{0.907} & \textbf{3.268} & 0.055 \\
 & single0017 & 0.869 & 0.740 & 3.025 & \textbf{0.032} \\
\midrule
\multirow{4}{*}{Happy}
 & Ceiling    & 0.933 & 0.853 & 2.984 & 0.058 \\
 & Base       & 0.425 & 0.951 & 3.391 & 0.056 \\
 & avg4spk    & \textbf{0.687} & \textbf{0.902} & \textbf{3.111} & \textbf{0.059} \\
 & single0017 & 0.686 & 0.850 & 3.104 & 0.065 \\
\midrule
\multirow{4}{*}{Sad}
 & Ceiling    & 0.953 & 0.914 & 3.235 & 0.053 \\
 & Base       & 0.540 & 0.948 & 3.397 & 0.059 \\
 & avg4spk    & 0.761 & \textbf{0.926} & \textbf{3.325} & \textbf{0.055} \\
 & single0017 & \textbf{0.816} & 0.841 & 2.913 & 0.086 \\
\bottomrule
\end{tabular}
\end{table}

\begin{figure}[H]
\centering
\includegraphics[height=4.6cm]{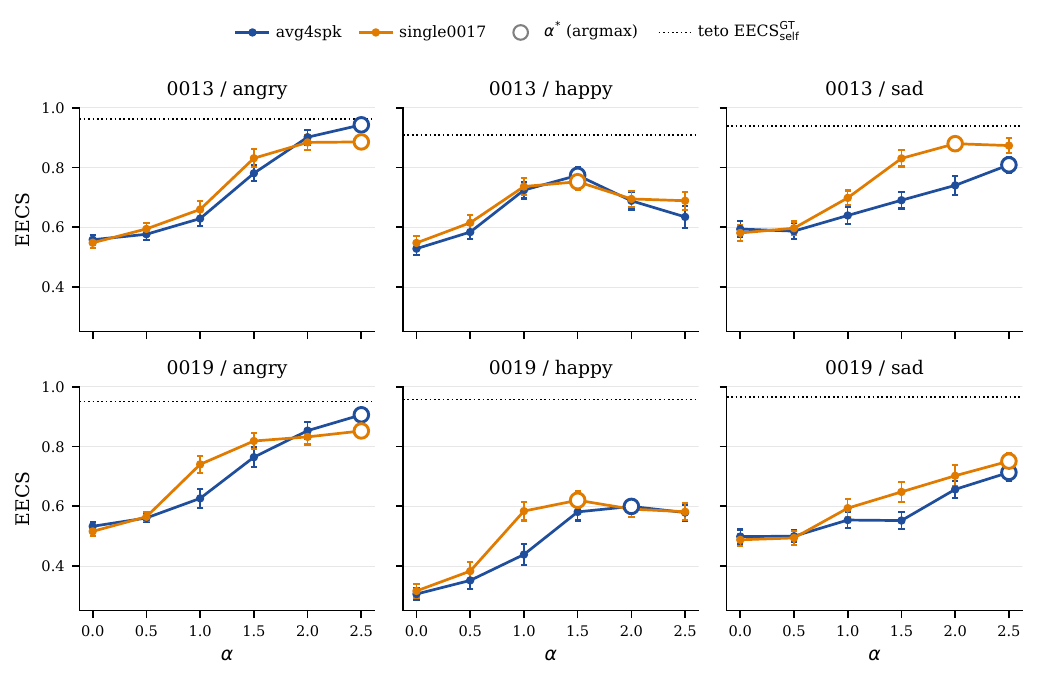}\hfill
\includegraphics[height=4.6cm]{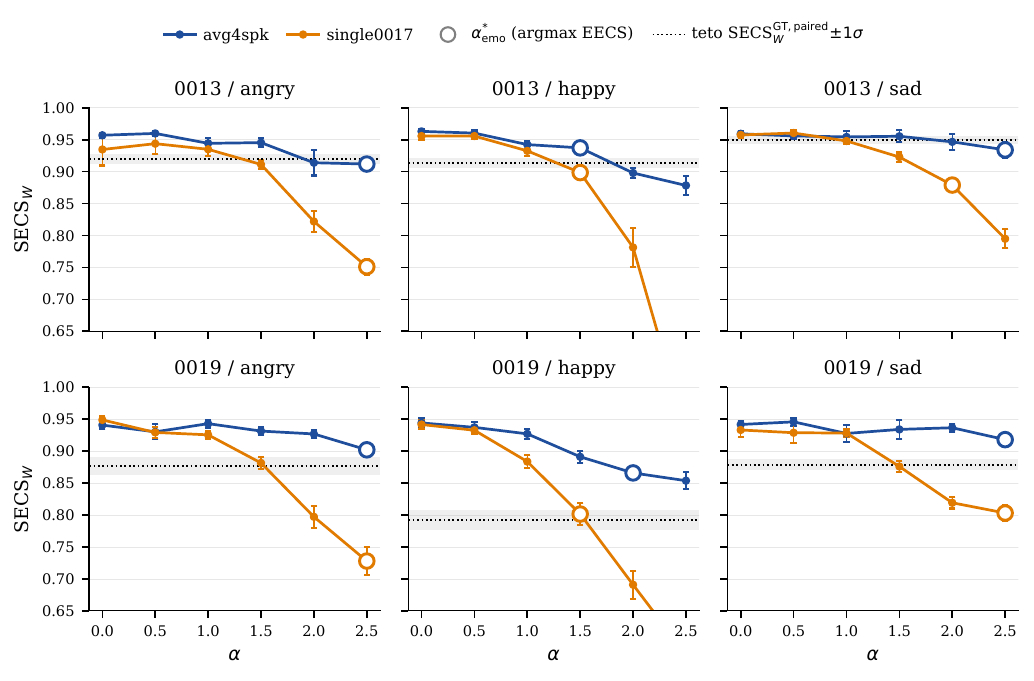}
\caption{$\alpha$ sweeps in the cross-speaker task arithmetic (EN held-out); \emph{avg4spk} (blue) and \emph{single0017} (orange), $\pm 1$ stderr ($n=30$), open marker at $\alpha^{*}$. \textbf{Left} (EECS): emotional transfer, with a dotted line at the natural ceiling $\text{EECS}_{\text{self}}^{\text{GT}}$. \textbf{Right} (SECS$_W$, independent WavLM encoder): identity preservation, band = ceiling $\text{SECS}_W^{\text{GT,paired}} \pm 1\sigma$; \emph{avg4spk} stays within or above the ceiling band, while \emph{single0017} falls below it at high $\alpha$ (identity$\times$intensity trade-off).}
\label{fig:en_sweep}
\end{figure}

Table~\ref{tab:en_operating} and Figure~\ref{fig:en_sweep} summarize the EN held-out results. The mean $\Delta$EECS gain over the pure ICL baseline is $+0.291$ (\emph{single0017}) and $+0.288$ (\emph{avg4spk}). The gains are positive across all six target$\times$emotion combinations, with $\Delta \in [+0.20, +0.39]$. WER$_{\text{norm}}$ remains stable ($\sim$5--7\%, identical to the human ceiling reachable by Whisper).

\textbf{Contrast $\tau_{\text{single}}$ vs.\ $\tau_{\text{avg}}$.} Aggregating the six combinations (target$\times$emotion) per variant, the multi-speaker average (an x-vector-space operationalization of the principle of \citet{murataSpeakeragnosticEmotionVector2025}) preserves the transfer gain ($\Delta$EECS $+0.288 \approx +0.291$), with a clear advantage in identity preservation ($+0.102$ points in SECS$_W$: $0.912$ vs.\ $0.810$) and in naturalness ($+0.20$ points in UTMOS: $3.23$ vs.\ $3.03$), despite a smaller norm (mean $\|\tau\|$ $1.60$ vs.\ $2.94$). The \emph{single0017} variant pays a toll in identity because it transports the source speaker's timbre residual to the target via $\tau$; averaging over four speakers cancels this idiosyncratic residual, preserving the shared emotional axis (cf.\ Table~\ref{tab:en_operating}).

\subsubsection{Cross-Lingual Validation with Ground Truth (emoUERJ PT-BR)}

\begin{figure}[!htbp]
\centering
\includegraphics[width=0.7\textwidth]{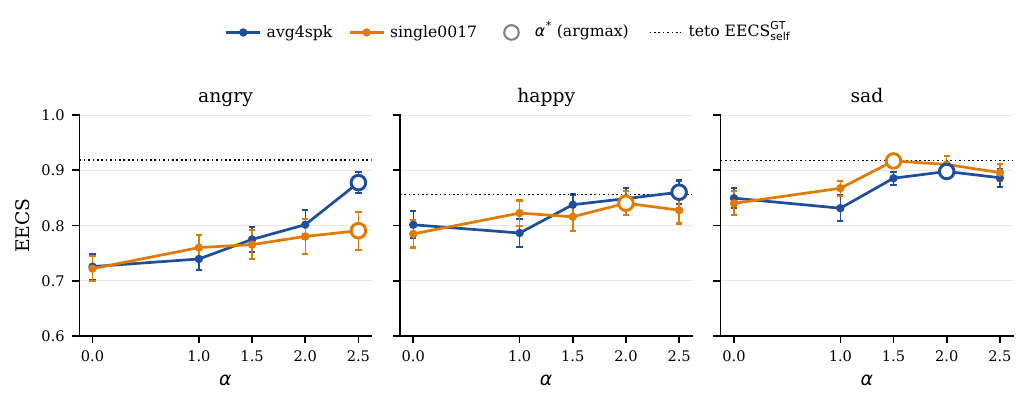}
\caption{$\alpha$ sweep (EECS) cross-lingual EN$\to$PT-BR (emoUERJ) with EECS aggregated over the three speakers \{m03, m04, w04\} by pooling the utterances; \emph{avg4spk} (blue) and \emph{single0017} (orange), $\pm 1$ stderr, open marker at $\alpha^{*}$; dotted line = natural ceiling $\text{EECS}_{\text{self}}^{\text{GT}}$.}
\label{fig:ptbr_sweep}
\end{figure}

Table~\ref{tab:ptbr_groundtruth} and Figure~\ref{fig:ptbr_sweep} summarize the cross-lingual EN$\to$PT-BR validation.

\begin{table}[!htbp]
\centering
\caption{Cross-lingual EN$\to$PT-BR validation (emoUERJ) at $\alpha^{*}_{\text{emo}}$ (mean over \{m03, m04, w04\}, $n \approx 6$--$14$ per combination). EECS = emotion2vec $\cos$ between synth and GT emo (paired for m03/m04, xtext for w04, whose SECS$_W$ is deflated by cross-text variance). Ceiling and Base as in Table~\ref{tab:en_operating}. UTMOS is not trained on PT-BR (relative within-experiment proxy), which depresses the human-ceiling UTMOS below the synthetic systems. Bold: best proposed variant per metric per emotion.}
\label{tab:ptbr_groundtruth}
\footnotesize
\setlength{\tabcolsep}{5pt}
\begin{tabular}{l l cccc}
\toprule
Emotion & System & EECS\,$\uparrow$ & SECS$_W$\,$\uparrow$ & UTMOS\,$\uparrow$ & WER$_{\text{norm}}$\,$\downarrow$ \\
\midrule
\multirow{4}{*}{Angry}
 & Ceiling    & 0.919 & 0.916 & 2.335 & 0.018 \\
 & Base       & 0.724 & 0.949 & 3.345 & 0.003 \\
 & avg4spk    & \textbf{0.877} & \textbf{0.929} & \textbf{3.162} & 0.000 \\
 & single0017 & 0.844 & 0.868 & 3.007 & 0.000 \\
\midrule
\multirow{4}{*}{Happy}
 & Ceiling    & 0.856 & 0.922 & 2.530 & 0.070 \\
 & Base       & 0.786 & 0.936 & 3.309 & 0.017 \\
 & avg4spk    & \textbf{0.867} & \textbf{0.902} & \textbf{2.996} & \textbf{0.000} \\
 & single0017 & 0.865 & 0.847 & 2.939 & 0.033 \\
\midrule
\multirow{4}{*}{Sad}
 & Ceiling    & 0.918 & 0.925 & 2.850 & 0.077 \\
 & Base       & 0.855 & 0.940 & 3.419 & 0.000 \\
 & avg4spk    & 0.902 & \textbf{0.939} & \textbf{3.267} & 0.002 \\
 & single0017 & \textbf{0.923} & 0.917 & 3.060 & \textbf{0.000} \\
\bottomrule
\end{tabular}
\end{table}

Most PT-BR combinations improve over the pure ICL baseline, with a modest mean gain ($\Delta = +0.092$); the gain is larger and clearer for high-arousal emotions (angry: $\Delta$ up to $+0.19$), yet smaller than in EN$\to$EN and small in some of the happy/sad combinations. This seemingly reduced gain stems largely from the starting point: the baseline (unmanipulated, $\alpha=0$) in PT-BR already reaches EECS $\in [0.70, 0.91]$, about $0.20$ above the EN$\to$EN baseline ($0.31$--$0.59$). Two causes contribute: the neutral speech of emoUERJ is less prototypically neutral (the neutral reference of m04, for example, sounds slightly tired), so the unmanipulated audio already carries some emotional coloring; and \texttt{emotion2vec\_plus\_large} is less discriminative in PT-BR, assigning high cosines even to barely emotional speech. Since the baseline starts from an already elevated level, there is less headroom for improvement, and the gain drops from $+0.29$ (EN) to $+0.09$ (PT-BR). In absolute value, however, the best $\alpha$ reaches EECS $\in [0.76, 0.97]$, the same range as EN$\to$EN: the final quality is equivalent, and only the \emph{gain} appears smaller.

In terms of identity preservation, the \emph{avg4spk} variant is substantially better in the cross-lingual setting: SECS$_W$ stays $\gtrsim 0.88$ across all nine \emph{avg4spk} combinations, whereas \emph{single0017} drops to $0.70$ in m03/happy ($\alpha=2.0$). WER remains $\approx 0$ across all combinations (mean $0.006$).

\begin{figure}[!htbp]
\centering
\includegraphics[height=5.6cm]{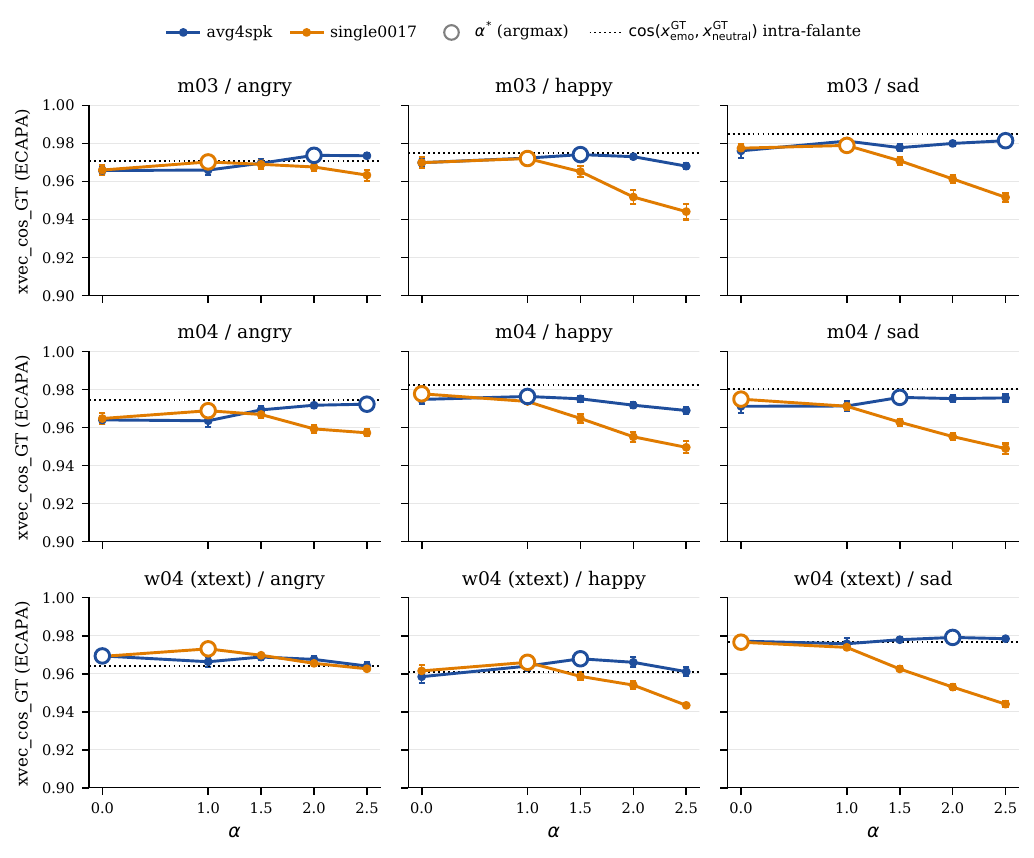}
\caption{Cross-lingual falsification test: $\overline{\text{xvec\_cos\_GT}}(\alpha)$, ECAPA-TDNN cosine between the synthesized x-vector and the GT emo recording of the same PT-BR speaker; $\tau_{\text{avg4spk}}$ (blue), $\tau_{\text{single0017}}$ (orange), $\pm 1$ stderr; dotted line = natural within-speaker shift.}
\label{fig:ptbr_curve}
\end{figure}

Figure~\ref{fig:ptbr_curve} complements this reading with a metric that is language-agnostic by construction (\texttt{xvec\_cos\_GT}, the cosine in ECAPA-TDNN space between two PT-BR audios of the same speaker). The metric saturates near the natural within-speaker limit, since the ECAPA space is dominated by identity and compresses prosodic variation, so it does not grow appreciably with $\alpha$. The informative signal lies in the contrast between variants: \emph{avg4spk} stays on the reference line throughout the entire $\alpha$ grid (the synthesized x-vector does not move away from the target identity), while \emph{single0017} falls below it at high $\alpha$, evidencing the source speaker's timbre leakage. This confirms that the English $\tau$ is language-agnostic and that the multi-speaker average transports the emotional axis without corrupting the PT-BR target's x-vector.

\section{Discussion}

The joint reading of Steps 1--4 and the PT-BR validation supports three methodological tensions that deserve further examination.

\subsection{Why Weight-Space Arithmetic Fails in LM-TTS}

The fundamental difference between modular architectures (CFS2, VITS) and LM-TTS lies in the functional locus of emotion among the stacked neural components. In CFS2, prosody is parameterized explicitly in pitch/energy/duration predictors (dedicated sub-networks); modifying the weights of these modules directly alters the conditioning. In VITS, \citet{kalyanEmotionArithmeticEmotional2024} localized emotion in the 1D convolutional layers of the text encoder, again a specific parameterized module. In Qwen3-TTS, these modules do not exist as a separate architectural block; prosody emerges from autoregressive continuation conditioned on the x-vector and the ICL tokens. Arithmetic on the backbone weights does not alter the conditioning, and single-speaker/single-emotion fine-tuning produces loss-quality divergence \citep{purwarWhenFineTuningFails2026}, without an intermediate regime of controlled emotional speech (Step~3, through controlled dissociation, points to the x-vector as the dominant carrier).

\subsection{The X-Vector as Emotion Carrier}

Traditionally, x-vectors are conceived as representations of vocal identity, and the speaker-verification literature treats within-speaker prosodic variation as noise to be discarded. In Qwen3-TTS, however, the ECAPA-TDNN speaker encoder is not a frozen pretrained verifier: it is \emph{co-trained} with the backbone under the synthesis objective \citep{huQwen3TTSTechnicalReport2026}, and the x-vector it produces is injected directly into the sequence of codec embeddings that conditions the LM, without intermediate projection. This configuration helps explain why emotion resides in this vector: since the Base model depends exclusively on the x-vector for zero-shot cloning, the reconstruction objective pressures the embedding to retain the prosodic-emotional cues necessary for expressive speech, not only identity. Our results, aligned with \citet{shaheenExploitingEmotionInformation2023, koharaSpeechEmotionControlTexttoSpeech2023}, empirically confirm this retention; indeed, the ICL baseline ($\alpha=0$) in PT-BR already reaches EECS $\in [0.70, 0.91]$ (Section~\ref{sec:experimentos}), consistent with an encoder that preserves, rather than discards, prosodic-emotional cues.

The ESD statistics are revealing: the norm of the emotional direction $\tau$ is only 15\% of the x-vector norm, and the projection of $\tau$ onto the identity axis is negligible (less than 1\% of $\|\tau\|$); the emotional direction is thus practically orthogonal to the identity axis. It is precisely this small magnitude that produces the near-collinearity between the neutral and angry x-vectors ($\cos(\mathbf{x}(0017,\text{neutral}), \mathbf{x}(0017,\text{angry})) = 0.988$). The x-vector remains a dominant identity bottleneck, but it carries, along this reduced-norm direction, an effective emotional axis. It is precisely in this component, whose approximately linear structure admits controlled interventions, that centroid arithmetic works, unlike the backbone weights and the discrete tokens.

\subsection{Identity--Intensity Trade-off and Metric Saturation}

Table~\ref{tab:en_operating} and Figure~\ref{fig:en_sweep} expose a trade-off between emotional transfer and identity preservation that $\alpha$ controls directly, without retraining. Decreasing $\alpha$ monotonically recovers identity at the cost of emotional fidelity; the operating point is selectable post-hoc over the $\alpha$ sweep, without re-synthesis. The \emph{avg4spk} variant mitigates the trade-off at its source: by canceling the source speaker's timbre residual, it preserves identity at $\alpha^{*}_{\text{emo}}$ in most PT-BR and EN combinations, whereas \emph{single0017} rarely does. This same trade-off was already observed in \citet{brito2026sintese}, but there the loss-weight control ($\alpha_{\text{speaker}}$ from 1 to 9 in VECL-TTS) did not perceptibly recover identity; here, by contrast, $\alpha$ traverses the trade-off continuously and selectably post-hoc, without retraining.

Additionally, Table~\ref{tab:ptbr_groundtruth} exposes a metric saturation: EECS saturates at $\sim$0.85--0.97 (especially in PT-BR, with its less prototypical neutral prosody), so that additional gains in this region may reflect over-projection rather than real transfer. The \texttt{xvec\_cos\_GT} saturates by construction in EN$\to$EN (the ECAPA encoder compresses within-speaker variation), becoming informative only in PT-BR (Figure~\ref{fig:ptbr_curve}).

\subsection{Positioning Relative to Contemporary Work}

Compared to fine-tuning-based methods in LM-TTS---IndexTTS2 with GRL+Perceiver Conditioner (three weeks on 8 A100 GPUs) \citep{zhouIndexTTS2BreakthroughEmotionally2025} and MiniMax-Speech with one LoRA per emotion \citep{zhangMiniMaxSpeechIntrinsicZeroShot2025}---our method gives up control sophistication (they offer natural-language instruction and native multilingualism) to gain in ergonomics: it dispenses with fine-tuning, can in principle operate over any LM-TTS with a learnable speaker encoder trained end-to-end for synthesis (although here validated only on Qwen3-TTS), and has a cost dominated by a single average of embeddings.

With respect to \citet{suniStyleProsodyControl2025}, the method is conceptually equivalent (training-free centroid arithmetic in embedding space; we differ only in the centroid estimation, by mean rather than geometric median). The contribution lies in offering empirical evidence that the gap they declare open, the incompatibility with token-based TTS, can be circumvented when the operand migrates from the dedicated style embedding (which LM-TTS does not have) to the learnable speaker embedding (which LM-TTS does have), even though the observed gains are modest in the cross-lingual setting. Additionally, we demonstrate cross-lingual EN$\to$PT-BR transfer with a parallel ground-truth reference, an aspect that \citet{suniStyleProsodyControl2025} do not explore.

An instructive contrast is the contemporary SelfTTS \citep{uedaSelfTTSCrossSpeaker2026}, which attacks the same cross-speaker style-transfer task for neutral-only speakers from the opposite direction: over a VITS architecture with \emph{learned} speaker and emotion encoders, it imposes \emph{explicit} disentanglement (a Gradient Reversal Layer with a cosine loss between the embeddings, together with a multi-positive contrastive loss), obtaining near-orthogonal emotion and speaker spaces (CKA $\approx 0.014$) by construction. Our finding is complementary and in the inverse direction: a speaker encoder co-trained end-to-end for synthesis appears to expose an approximately linear emotional direction in x-vector space \emph{without} any dedicated disentanglement loss, enabling the training-free arithmetic. In other words, where SelfTTS \emph{engineers} the identity--emotion separation, we \emph{exploit} it where it emerges from co-training---two converging pieces of evidence that the separability of these representations is what enables cross-speaker transfer.

\subsection{Limitations}

Five caveats delimit the scope. \textbf{(i)~Generalization:} the experiments are restricted to Qwen3-TTS-12Hz-1.7B; the architectural argument applies to any system with ICL and a learnable speaker encoder co-trained with the backbone, but validation on other models remains to be done. \textbf{(ii)~Encoder dependence:} the method assumes that the speaker encoder captures emotional prosody in approximately linear directions; encoders trained exclusively for speaker verification, or with explicit identity--emotion disentanglement (e.g., GRL), may compress the operational window of $\alpha$. \textbf{(iii)~PT-BR metrics:} \texttt{emotion2vec\_plus\_large} was trained mostly on ZH/EN, contributing to the observed saturation---mitigated by \texttt{xvec\_cos\_GT}, language-agnostic by construction, with a human listening test pending. \textbf{(iv)~Explicit conditioning:} variants trained with instruction--speech pairs (CustomVoice, VoiceEditing) offer more granular control and are preferable when they exist; our contribution is complementary, aimed at the Base model (purely ICL), common among open research LM-TTS. \textbf{(v)~Step-1 data regime:} the negative result of weight-space arithmetic rests on single-speaker/single-emotion fine-tuning ($\sim$30 min of speaker 0017), a low-acoustic-variability regime in which \citet{purwarWhenFineTuningFails2026} report loss-quality divergence. We therefore do not rule out that larger-scale multi-speaker/multi-emotion fine-tuning would open an operational window; the localization of emotion in the x-vector, however, rests chiefly on Step 3 (\texttt{full\_swap}), independent of the fine-tuning regime.

\section{Conclusions}

This work investigated whether task-vector arithmetic, successful in modular TTS, transfers to modern LM-TTS systems based on language-model backbones with in-context learning. The answer is negative when the arithmetic operand is the backbone weights, as Step 1 indicates that, in the single-speaker/single-emotion regime we tested, fine-tuning does not produce controllable emotional speech within a useful operational window. Through progressive elimination over four candidate operands in stacked neural networks, we localize the dominant carrier of emotion in the speaker embedding extracted by the ECAPA-TDNN encoder. Building on this localization, we propose a training-free method of multi-speaker centroid arithmetic in x-vector space and demonstrate cross-speaker (EN held-out: $\Delta$EECS $+0.29$ over the baseline) and cross-lingual EN$\to$PT-BR ($\Delta$EECS $+0.09$, modest but positive in most combinations) transfer, with emotional intensity controllable by a single inference hyperparameter. These results suggest that the incompatibility of centroid arithmetic with token-based TTS architectures can be circumvented by operating on the speaker embedding, offering a low-compute alternative to contemporary methods in emotional LM-TTS that rely on dedicated fine-tuning.

As future work, we intend to replicate the method on other open LM-TTS, compose emotions via linear combination of $\tau$, normalize $\alpha$ by cosine distance between speakers, and conduct a human listening test in PT-BR.

\bibliographystyle{unsrtnat}
\bibliography{references}

\end{document}